\documentclass[preprint,aps]{revtex4}

\usepackage{graphicx}
\usepackage{dcolumn}
\usepackage{bm}

\begin{document}


\title{Raman microspectroscopic characterization of amorphous silica plastic behavior}

\author{A. Perriot$^{1}$, V. Martinez$^{2}$, L. Grosvalet$^{2}$, Ch. Martinet$^{2}$, \linebreak B. Champagnon$^{2}$, D. Vandembroucq$^{1*}$ and E. Barthel$^{1}$}
\email{damien.vdb@saint-gobain.com}

\affiliation{
$^{1}$ Surface du Verre et Interfaces, UMR 125 CNRS/Saint-Gobain, Aubervilliers, France.\\
$^{2}$Laboratoire de Physico-Chimie des Mat\'eriaux Luminescents,
UMR 5620 CNRS/Universit\'e Claude Bernard---Lyon I, Villeurbanne,
France.}

\date{\today}

\begin{abstract}
Raman microspectroscopy was used to characterize amorphous
silica plastic behavior. Using a correlation between Raman
spectrum and density, a map of the local residual
indentation-induced densification is obtained. The existence of a
densification-induced hardening is also evidenced through a
diamond anvil cell experiment. Such observations are not accounted
for by the previously proposed hardening-free pressure-dependent
yield criterion based on indentation curves. These results open
the way to more accurate description of a constitutive law for
amorphous silica.
\end{abstract}

\maketitle


\section{Introduction}
Brittleness is a well-known and extensively studied characteristic
of silicate glasses \cite{zar}. Comparatively, their plastic
behavior has remained far less explored. Yet, the latter
determines the contact mechanics of the glass surface
\cite{griffith} which is of primary interest for many micron-scale
applications.

For obvious historical reasons, the standard vision of plasticity
is still largely focused on crystalline materials (especially
metals), where irreversible strain is triggered by shear stress
and accommodated through volume-conserving dislocations motion
\cite{cotrell}. Glasses do not fit into this frame for two
different reasons. First, because of their amorphous nature there
is no direct equivalent to these dislocation-based mechanisms
\cite{zar}. Second, being amorphous, their structures,
significantly more open than those of the chemically equivalent
crystals \cite{zar}, makes them prone to densify. Indeed, silicate
glasses densify \cite{Ernsb} while experiencing a plastic behavior
that can be activated even by a purely hydrostatic loading
\cite{sug1}. This last characteristic leads to a distinction
between normal and anomalous glasses, which exhibit respectively
little and large densification, as well as different indentation
fracture patterns \cite{hagan,hagan2}. For fused silica (density
2.2 g.cm$^{-3}$), the most anomalous silicate glass of industrial
interest, this densification process is far from being negligible
\cite{kurkjian} . It is acknowledged to densify up to about 20
$\%$ under mechanical stress \cite{lambini}. Although this
anomalous behavior of fused silica has been known for half a
century \cite{taylor}, in-depth investigations were limited by the
brittle nature of silica, which restricts the characteristic size
of plastic deformation to micrometer scale \cite{rhee}. Indeed,
evidences of the densification process at room-temperature can
only be obtained experimentally in confined geometries, such as in
a Diamond-Anvil Cell \cite{sug1} (DAC) or under an indenter
\cite{taylor,Ernsb}, which do not allow direct access to local
stress/strain data through the commonly used macroscopic
techniques.

We show here that, for amorphous silica, such local data of
primary importance can be obtained experimentally through Raman
microspectroscopy, which has recently proven particularly
interesting as a method for local material characterization
\cite{ramanmicro,fonzo}. Through the combination of a specific
method to obtain unaltered cross-sections of indentations with
Raman microspectroscopic mappings, we provide a map of the local
indentation-induced densification in a fused silica sample. This
experimental data on the local strain state of the material
constitutes a relevant sieve for the validation of a constitutive
law for amorphous silica
\cite{imaoka,imaoka2,lambini,lamb2d,lamb3d,lambbis,lambter}.

The most sophisticated constitutive law for amorphous
silica \cite{lamb3d} does not account for these results. We
suggest it be refined further by taking into account amorphous
silica densification-induced hardening, which we also evidence
here with a DAC experiment.

\section{Experimental procedure}

\subsection{Using Raman spectrum as a density gauge}

\begin{figure}
\includegraphics[width=8 cm]{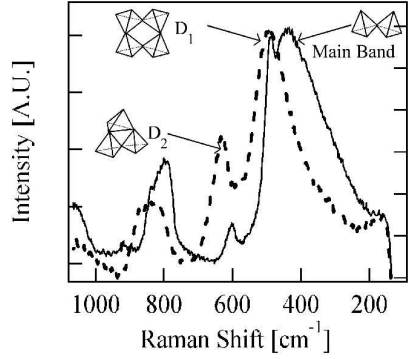}
 \caption{Raman spectra for amorphous
silica. The plain line spectrum was obtained with 2.2 g.cm$^{-3}$
amorphous silica, while the dotted line corresponds to
indentation-densified amorphous silica.}\label{spectre}
\end{figure}

Between 200 and 750 cm$^{-1}$, the Raman spectrum of silica glass
consists in three bands (Fig. \ref{spectre} contrasts the Raman
spectrum of non-densified fused silica with the spectrum of
largely indentation-densified silica glass). The main band at 440
cm$^{-1}$ results from the symmetric stretching mode of bridging
oxygens between two Si atoms. This band is intense, and affected
by the densification process through a decrease of the
inter-tetrahedral angles Si-O-Si
\cite{Tomozawa_uniaxial,mochizuki}, but it was shown to be also
very sensitive to elastic loading \cite{Mich} (such as the
residual elastic strain expected in indentation).

The defect lines D$_1$ and D$_2$, at 492 and 605 cm$^{-1}$, are
respectively attributed to the breathing modes of the
four-membered and three-membered rings \cite{galeener}. Their
intensity ratio was previously used in literature as an indicator
for ``fictive temperature'' \cite{champi,champii,champiii}, a
slight change of which corresponds to a slight change in density.
Moreover, they shift in position towards higher wavenumbers with
increasing density, as the associated breathing modes get
frustrated by the lack of free volume.

The D$_2$ line is particularly useful because it has almost no
overlap with the main band. Sugiura et al. \cite{sug2} correlated
the position of the D$_2$ line with the ratio of the sample
density $\rho$ to its initial density $\rho_0$. From their work,
we get the following empirical relation:
\begin{equation}\label{indicator}
0.143\,\log_{10}\left(\frac{\rho}{\rho_0}\right)=\log_{10}\left(\frac{\nu}{\nu_0}\right)
\end{equation}
$\nu$ and $\nu_0$  being respectively the final and the initial
band position. The residual density evaluated through this
relation accounts for both irreversible and elastic densification
due to residual elastic strain. However, several works
\cite{Mich,sug2} evidenced that the D$_2$ line position is only
marginally sensitive to residual elastic strains.

\subsection{Diamond-Anvil Cell geometry}

An amorphous silica (Saint-Gobain Quartz IDD) splinter with
a ten-micron characteristic length was used as experimental
sample.
It was introduced into a DAC apparatus 
along with a very small piece of ruby. A Raman spectrum of the
sample was acquired prior to the filling-up of the cell
compartment with the pressure transmitting fluid. We used a
transmitting fluid close to 1:5 methanol-ethanol mix, which
insures hydrostaticity up to 16 to 20 GPa \cite{metheth}. That
choice was motivated by the fact that the densification process in
amorphous silica is expected to saturate between 20 and 25 GPa
under hydrostatic loading \cite{polgrimsat}.

The sample was then loaded to a maximal applied hydrostatic
pressure evaluated to 14 GPa, using the shift of ruby fluorescence
bands. After complete unloading, the Raman spectrum of the sample
was measured.

\subsection{Indentation geometry}

Samples were prepared according to the following procedure. Plates
of amorphous silica (IDD, Saint-Gobain Quartz), of dimension
10$\times$10$\times$0.4 cm$^3$, were indented with an instrumented
microindentation device allowing surface visualization. These
plain 2-kg-Vickers indents allowed us to gain access to top views
of the densified area.

2-kg-Vickers indents were also performed on a previously
subcritically-grown crack, as described on figure \ref{protocole
tranche}. This provided us, on the crack surface, with a
cross-section view of the densified material area located under
the indent. In order to minimize the mechanical perturbation due
to the presence of the crack, we performed a symmetrical loading.
The indenter was thus positioned so that one of the Vickers
pyramid diagonals be superimposed with the crack. Indentation
curves obtained for on-crack indentations were similar
(figure \ref{indentation_curves} shows that both
indentation curves almost coincide) with that obtained for plain
2-kg-Vickers indentations, showing that the presence of a crack
did not noticeably affect the indentation process.

We used a Raman microspectroscopy device (RM1000, Renishaw)
including a 30 mW \linebreak 514 nm Argon laser, under a
$\times$100 objective. The laser beam was focused on the surface
without confocal system. We however checked on one section view
that the mapping obtained using confocal system agreed with that
obtained without confocal system. We evaluated the excited volume
of amorphous silica to approximately 2$\times$2$\times$5
$\mathrm{\mu m}^3$. The characteristic size of that scanned
volume, which directly depends on the laser wavelength, is too
large to map an elastoplastic nanoindent. However, it can easily
be used on microindents, such as that obtained with 2-kg-Vickers
indents. Under the assumption that microcracks have no significant
effect on the indentation-induced strain field, such experimental
data can be used to validate an elastoplastic constitutive law for
amorphous silica.

After checking densification symmetry with respect to the indenter
geometry, we performed surface mappings. Owing to the symmetry of
the Vickers indenter, we first mapped one eighth of the densified
area on three different plain indents. We thus recorded spectra in
13 locations with a 10 $\mathrm{\mu m}$-spacing, according to Fig.
\ref{cartobis}, during 1500 s each.

We then mapped half of the densified area on three different
section views, recording spectra in 42 locations with a 5
$\mathrm{\mu m}$-spacing, according to Fig. \ref{carto}. We also
performed similar measurements on subsurface positions in order to
complete the section map. These measurements are to be compared
with those obtained under indent edges on top view mappings. For
each spectrum, Raman scattering signal was accumulated during 900
s.

\section{Results}

\subsection{Diamond-anvil cell geometry}

Figure \ref{spectra} contrasts the initial spectrum of the
sample (plain line) and that obtained after a 14 GPa hydrostatic
loading (dotted bold line). Using relation (\ref{indicator})
between the D$_2$ line shift and densification, the densification
experienced by the sample is derived directly from the spectra
measured before and after hydrostatic loading. Thus the sample
experienced a densification of approximately 10$\%$.

\subsection{Indentation geometry}

The D$_2$ peak shift was then used as a density estimator
according to equation (\ref{indicator}). A typical top view map is
displayed on Fig. \ref{cartobis}. A typical cross-section map is
displayed on Fig. \ref{carto}.

The iso-densification lines obtained for the top view map (Fig.
\ref{cartobis}) are concentric. Their shape does account for the
anisotropy of the Vickers pyramid. Indeed, densification is more
pronounced under the indenter edge than it is under the indenter
face. This result is consistent with the fact that shear stress
(necessarily more intense under acuter parts of the Vickers
pyramid) enhances densification \cite{Mck63}.

The iso-densification lines obtained for the cross-section map
(Fig. \ref{carto}) are shaped into ``concentric bowls'', which
agrees with the optical observations by Hagan and Van der Zwaag
\cite{hagan}. Quantitatively, we evaluated the densification just
under the indenter tip at around 16 $\%$. These results are
consistent with the saturation of the densification process that
is accepted to occur around 20$\%$ \cite{lambini}. Moreover, we
notice that the subsurface measurements are consistent with those
obtained along the indent diagonal on top view maps. This
validates that the indents performed on pre-existing cracks are
indeed reasonably equivalent to plain indents. The densification
just under the indenter tip was again estimated to approximately
16 $\%$, in agreement with the previous measurement.

\section{Discussion}

\subsection{Yield criterion for amorphous silica}

In most crystalline materials, plasticity occurs when the shear
component of the applied stress reaches a yield value. Formally,
this yield criterion can be written :
\begin{equation}\label{mises}
f(\underline{\underline{\sigma}})=\tau-Y
\end{equation}
where $\tau$ represents the intensity of shear stress and $Y$ the
yield strength. As long as $f(\underline{\underline{\sigma}})$
remains negative (i.e. as long as the shear stress remains smaller
than the material yield strength), the material behaves
elastically, while it may behave plastically only when\linebreak
$f(\underline{\underline{\sigma}})=0$. However, such a criterion
cannot account for the silica glass densification under purely
hydrostatic loading \cite{sug1}. In order to do so, a dependance
of $f(\underline{\underline{\sigma}})$ to the hydrostatic pressure
$p$ has to be taken into account.

Recently, a hydrostatic-pressure-dependent yield criterion was
proposed to describe the plastic behavior of dense amorphous
materials like silica glass
\cite{lambini,lamb2d,lamb3d,lambbis,lambter}. This plastic yield
criterion, $f(\underline{\underline{\sigma}})$, can be written:
\begin{equation}\label{criterion}
f(\underline{\underline{\sigma}})=\alpha
p+(1-\alpha)\tau-Y=-\frac{\alpha}{3}Tr(\underline{\underline{\sigma}})+(1-\alpha)\sqrt{\frac{\underline{\underline{s}}:\underline{\underline{s}}}{2}}-Y
\end{equation}
where
$Tr(\underline{\underline{\sigma}})=\sigma_{11}+\sigma_{22}+\sigma_{33}$
and
$\underline{\underline{s}}=\underline{\underline{\sigma}}-Tr(\underline{\underline{\sigma}})\underline{\underline{I}}$
is the deviatoric part of $\underline{\underline{\sigma}}$.

This pressure-dependent yield criterion is parameterized by two
physical characteristics. One of them is the yield strength, $Y$.
The other one, known as the densification factor \cite{imaoka}
$\alpha$, balances the respective influence of applied hydrostatic
pressure, $p$, and shear stress, $\tau$.
 For $\alpha = 0$, we obtain the Van Mises criterion (Eq.
(\ref{mises})), which describes the densification-free plasticity
of metals and, for $\alpha = 1$, we describe a densification-only
plastic behavior triggered exclusively by hydrostatic pressure,
provided the plastic behavior is associated. Hence the
denomination of $\alpha$ as the densification factor: the greater
$\alpha$, the more anomalous the material.

Under the assumption of an associated perfect-plastic behavior,
Xin and Lambropoulos \cite{lamb3d} adjusted these two parameters
by comparing experimental load/displacement indentation curve with
finite element simulations. With the values $\alpha = 0.6$ and $Y
= 5.43$ GPa, they reproduced their experimental data: $\alpha =
0.6$ accounts for a large contribution of densification to plastic
strain (even allowing fused silica to exhibit a slight contraction
of the cross section while being uniaxially compressed); and the
hydrostatic loading yield pressure, obtained for $\tau = 0$, is
consistent with DAC experimental results \cite{sug2}.

\subsection{Amorphous silica densification-induced hardening}


We provide here a simple argument proving that amorphous
silica hardens. Let us assume amorphous silica has a perfect
plastic behavior (that is no hardening), at least until it reaches
saturation density (See figure \ref{perfect}). When
hydrostatically loaded, before saturation occurs, the applied
stress on amorphous silica will never exceed the material yield
strength, which is the only applied stress at which plastic flow
occurs in a perfect plastic material. Thus, amorphous silica has
necessarily reached saturation density before it can reach an
applied pressure higher than its yield stress. Our DAC sample was
loaded above 9 GPa, which is the yield pressure $p^Y_0$ provided
by Xin and Lambropoulos' model. However, its densification is
evaluated to a mere 10\%, which is lower than both literature
claims (20\%) and our measurements under the indenter tip (16\%)
for the saturation density. This leads to the conclusion that
amorphous silica indeed exhibits hardening. Moreover, similar DAC
cycling-loading experiments would provide us with the evolution of
amorphous silica yield criterion with material density along the
hydrostatic pressure axis.

\subsection{Densification map: a relevant local characterization of indentation-induced strain field}

Indentation experiments result in largely inhomogeneous
stress distributions. The applied force (resp. the penetration) is
merely the integral of local applied pressure (resp. vertical
displacement) and as such simply proportional to the mean applied
pressure (resp. vertical displacement). The exclusive use of such
integral data may prove insufficient to identify the parameters of
a constitutive law trying to account for more local strain
measurements. Our Raman microspectroscopic maps may constitute a
severe validation for such laws and it may help to gain insight on
the refinements to perform.

To test the influence of the densification factor on the general
shape of the residual densified area, we performed FE
calculations, using the software Abaqus, to model a cone
indentation (of semi-apical angle of 70.3$^o$ --- making it the
``axisymmetric equivalent'' to a Vickers indenter). Consistently
with Xin and Lambropoulos' approach \cite{lamb3d}, the material
plastic behavior was assumed perfect and associated. The
saturation, accounted for by preventing the material from
densifying or flowing any longer, occurs when material permanent
densification reaches 16$\%$. Calculations were performed for
several values of $\alpha$ ranging between 0.2 and 0.6. The yield
strength, $Y$, was chosen to match the experimental value in
hydrostatic conditions ($Y/\alpha = 9.05$ GPa). Residual
densification maps for $\alpha = 0.2$ and $\alpha= 0.6$ are
presented in Fig. \ref{densif}. When comparing these two maps, we
notice that the indentation-densified areas are markedly
different. In the case $\alpha = 0.6$, with a large dependence on
hydrostatic contribution, the numerical results predict a small
but almost entirely saturated densified area . However, in the
less anomalous case $\alpha = 0.2$, the densification area is
larger but with a smoother transition towards the non-densified
material. This contrast highlights the direct relevance of the
density map for the identification of the constitutive law.

\subsection{The role of strain-hardening in amorphous silica
plasticity}

Let us now compare our experimental density map with the numerical
results proposed by the literature. The densified area obtained by
Xin and Lambropoulos \cite{lamb3d} with $\alpha = 0.6$ and $Y =
5.43$ GPa is qualitatively close to that provided by figure
\ref{densif}(b). The general shape and extension of the densified
area is recovered. This confirms that the densification
contribution to the silica plastic behavior is significant
\cite{lambini,lamb2d,lamb3d}. However, the mainly saturated core
with a sharp near-boundary density gradient largely contrasts with
the more progressive densification of the experimental result.
This shows clearly that the constitutive law for silica
densification identified using indentation curves needs to be
further refined to account for the indentation-induced
densification map.

At least one of the assumptions made in the constitutive law has
to be relaxed. We expect that taking into account the
densification-hardening process, the existence of which we have
demonstrated by DAC measurements, will result in the concentric
``bowl-shaped'' areas associated to the experimental smoother
density gradient, without much change in the global dimensions of
the densified area.

\section{Conclusion}

Raman microspectroscopy provides experimentally relevant
local data on the indentation-induced densification in amorphous
silica. They constitute a discriminant sieve for the
identification of its constitutive law, as they cannot be
accounted for by existing models identified through indentation
curves. Our results also clearly evidence a densification-induced
hardening process. Taking it into account would allow one to model
more accurately the plastic behavior of amorphous silica.

We are currently working on finite-element simulation to
provide such a plastic criterion for amorphous silica behavior.
On-going experimental work also investigates the use of
microspectroscopic methods to provide useful information on the
mechanical behavior of other more complex silicate glasses.

\section*{Acknowledgments}

The authors would like to thank G. Duisit, R. Gy, S. Pelletier, G.
Qu\'erel and S. Roux for their help, advice and support.

\bibliographystyle{unsrt}
\bibliography{these}

\begin{thebibliography}{10}

\bibitem{zar}
J. Zarzycki,
\newblock {\em Glasses and the vitreous state};
\newblock Cambridge University Press, 1991.

\bibitem{griffith}
A. Griffith,
\newblock {\em Philos. Trans. Roy. Soc.}, A221 163 (1920).

\bibitem{cotrell}
A.H. Cotrell,
\newblock {\em Dislocations and Plastic Flow in Crystals};
\newblock Clarendon Press, Oxford, 1953.

\bibitem{Ernsb}
F.M. Ernsberger,
\newblock ``Role of densification in deformation of glasses under point loading'',
\newblock {\em J. Am. Ceram. Soc.}, 51 [10] 545-547  (1968).

\bibitem{sug1}
H. Sugiura and T. Yamadaya,
\newblock ``Raman scattering in silica glass in the permanent densification
  region'',
\newblock {\em J. Non-Cryst. Solids}, 144 151-158 (1992).

\bibitem{hagan}
J.T. Hagan and S. Van Der Zwaag,
\newblock ``Plastic processes in a range of soda-lime-silica glasses'',
\newblock {\em J. Non-Cryst. Solids}, 64 249-268  (1984).

\bibitem{hagan2}
J.T. Hagan,
\newblock ``Cone cracks around vickers indentations in fused silica glass'',
\newblock {\em J. Mater. Sci.}, 14 462-466 (1979).

\bibitem{kurkjian}
C.R. Kurkjian and G.W. Kammlott,
\newblock ``Indentation behavior of soda-lime silica glass, fused silica, and
  single-crystal quartz at liquid nitrogen temperature'',
\newblock {\em J. Am. Ceram. Soc.}, 78 [3] 737-744, (1995).

\bibitem{lambini}
J.C. Lambropoulos, S. Xu, and T. Fang,
\newblock ``Constitutive law for the densification of fused silica, with
  applications in polishing and microgrinding'',
\newblock {\em J. Am. Ceram. Soc.}, 79 [6] 1441-1452  (1996).

\bibitem{taylor}
E.W. Taylor,
\newblock ``Plastic deformation of optical glasses'',
\newblock {\em Nature}, 163 323 (1949).

\bibitem{rhee}
Y.W. Rhee, H.W. Kim, Y. Deng, and B.R. Lawn,
\newblock ``Brittle fracture versus quasi plasticity in ceramics: A simple
  predictive index'',
\newblock {\em J. Am. Ceram. Soc.}, 84 [3] 561-565 (2001).

\bibitem{ramanmicro}
A. Kailer, K.G. Nickel and Y.G. Gogotsi,
\newblock ``Raman microspectroscopy of nanocrystalline and amorphous phases in
  hardness indentations'',
\newblock {\em J. Raman Spectrosc.}, 30 939-946 (1999).

\bibitem{fonzo}
S. DiFonzo, W. Jark, S. Lagomarsino, C. Giannini, L. DeCaro, A. Cedolla and M. Muller,
\newblock ``Non-destructive determination of local strain with 100-nm spatial
  resolution'',
\newblock {\em Nature}, 403 638-640 (2000).

\bibitem{lamb2d}
A. Shorey, K. Xin, K.H. Chen and J.C. Lambropoulos,
\newblock ``Deformation of fused silica : Nanoindentation and densification'',
\newblock {\em Proc SPIE}, 3424 72-81 (1998).

\bibitem{lamb3d}
K. Xin and J.C. Lambropoulos,
\newblock ``Densification of fused silica : Effects on nanoindentation'',
\newblock {\em Proc. SPIE}, 4102 112-121 (2000).

\bibitem{imaoka}
M. Imaoka and I. Yasui,
\newblock ``Finite element analysis of indentation on glass'',
\newblock {\em J. Non-Cryst. Solids}, 22 315-329 (1976).

\bibitem{imaoka2}
I. Yasui and M. Imaoka,
\newblock ``Finite element analysis of indentation on glass (ii)'',
\newblock {\em J. Non-Cryst. Solids}, 50 219-232 (1982).

\bibitem{lambbis}
F. Dahmani, J.C. Lambropoulos, A.W. Schmid, S.J. Burns and C. Pratt,
\newblock ``Nanoindentation technique for measuring residual stress field around laser-induced crack in fused silica'',
{\em J. Mater. Sci.}, 33 4677-4685 (1998).

\bibitem{lambter}
K. Xin and J.C. Lambropoulos,
\newblock ``Spherical cavity expansion in densifying material'',
{\em J. Appl. Phys.}, 94 6437-6441 (2003).

\bibitem{Tomozawa_uniaxial}
M. Tomozawa, Y.K. Lee and Y.L. Peng,
\newblock ``Effect of uniaxial stresses on silica glass structure investigated by
  ir spectroscopy'',
\newblock {\em J. Non-Cryst. Solids}, 242 [2-3] 104-109 (1998).

\bibitem{mochizuki}
S. Mochizuki and N. Kawai,
\newblock ``Lattice vibrationnal spectra of vitreous silica densified by
  pressure'',
\newblock {\em Solid State Communications}, 11 [6] 763-765 (1972).

\bibitem{Mich}
T.A. Michalske, D. Tallant and W.L. Smith,
\newblock ``Raman study of silica glass under tensile stress'',
\newblock {\em Phys. Chem. Glasses}, 29[4] 150-153 (1988).

\bibitem{galeener}
F.L. Galeneer,
\newblock ``Planar rings in glasses'',
\newblock {\em Solid State Communications}, 44[7] 1037-1040 (1982).

\bibitem{champi}
B. Champagnon, C. Chemarin, E. Duval and R. Le Parc,
\newblock ``Glass structure and light scattering'',
\newblock {\em J. Non-Cryst. Solids}, 274 81-86 (2000).

\bibitem{champii}
R. Le Parc, B. Champagnon, Ph. Guenot and S. Dubois,
\newblock ``Thermal annealing and density fluctuations in silica glass'',
\newblock {\em J. Non-Cryst. Solids}, 293-295 367-369 (2001).

\bibitem{champiii}
B. Champagnon, R. Le Parc and Ph. Guenot,
\newblock ``Relaxation of silica above and below T$_g$ : light scattering
  studies'',
\newblock {\em Phil. Mag. B}, 82 [2] 251-255 (2002).

\bibitem{sug2}
H. Sugiura, R. Ikeda, K. Kondo and T. Yamadaya,
\newblock ``Densified silica after shock compression'',
\newblock {\em J. Appl. Phys.}, 81 [4] 1651-1655 (1997).

\bibitem{metheth}
J.M. Besson and P.P. Pinceaux,
\newblock ``Uniform stress conditions in the diamond anvil cell at 200 kilobars'',
\newblock {\em Rev. Sci. Instrum.}, 50 [5] 541-543 (1979).

\bibitem{polgrimsat}
A. Polian and M. Grimsditch,
\newblock ``Room-temperature densification of a-SiO$_2$ versus pressure'',
\newblock {\em Phys. Rev. B}, 41 [9] 6086-6087 (1990).

\bibitem{Mck63}
J.D. Mackenzie,
\newblock ``High-pressure effects on oxide glasses'',
\newblock {\em J. Am. Ceram. Soc.}, 46 [10] 461-476 (1963).

\end{thebibliography}

\newpage

\section*{Figure Captions}

\underline{\textbf{Figure 1:}} Raman spectra for amorphous silica.
The plain line spectrum was obtained with 2.2 g.cm$^{-3}$
amorphous silica, while the dotted line corresponds to
indentation-densified amorphous silica.


\underline{\textbf{Figure 2:}} Experimental procedure allowing the
obtention of a section view of the indented area. Using a scratch
as surface defect (a), a subcritical crack is grown in the sample,
using a double-bending loading geometry (b). Symmetrical
indentation is then performed on the crack (c). Crack propagation
is then completed under double-bending (d). Two twin sections of
the indented area are then obtained.

\underline{\textbf{Figure 3:}} Indentation curves obtained
for both on-crack indentation and plain indentation. They almost
superimpose which shows us that the presence of a crack did not
significantly affect the indentation process.

\underline{\textbf{Figure 4:}} Densification map of one eighth of
a top view of an indentation-densified area obtained using  the
D$_2$ band position indicator. The center of the indent is at
coordinates (0;0). The bottom thin line figures the edge of the
indent while the right thin line corresponds to the middle of the
pyramid face. The crosses correspond to the measurement points.
The bold lines are the iso-densification lines. Numerical values
are given in percent. Iso-densification lines appear to be
concentric and account for the indenter anisotropy, as
densification appears more important under the indenter edge.

\underline{\textbf{Figure 5:}} Densification map of half a section
view of an indentation-densified area obtained using the D$_2$
band position indicator. The center of the indent is at
coordinates (0;0). The top bold line represents the sample
surface. The crosses correspond to the measurement points. The
black lines are the iso-densification lines. Numerical values are
given in percent. Iso-densification lines appear to be shaped into
concentric bowls in agreement with observations by Hagan et al.
\cite{hagan}

\underline{\textbf{Figure 6:}} Raman spectrum obtained with an
amorphous silica sample before (plain line) and after (dotted
line) a 14 GPa hydrostatic loading.

\underline{\textbf{Figure 7:}}Expected behavior of a
perfect plastic material under hydrostatic loading. Before
pressure reaches the yield pressure ($p^Y_0=9\mathrm{GPa}$) the
sample does not densify. As its plastic behavior is
hardening-free, applied pressure cannot reach higher values than
$p^Y_0$ until irreversible densification, $\Delta V/V$, reaches
its saturation value $\Delta V/V |_{sat}$.

\underline{\textbf{Figure 8:}} Residual densification maps
obtained through FE simulation of a 2D-axisymmetric cone
indentation. They were obtained for two different values of the
densification factor, (a) $\alpha = 0.2$ and (b) $\alpha = 0.6$.
Densification values are given in percent. The results by Xin and
Lambropoulos\cite{lamb3d} are similar to (b)

\newpage

\begin{figure}
\centering
\includegraphics [width=8 cm]{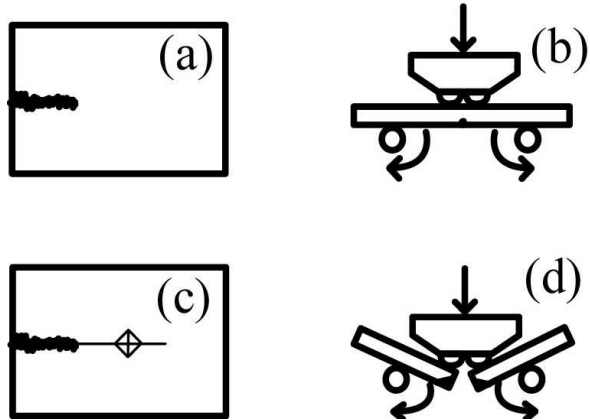}
\caption{Experimental procedure allowing the obtention of a
section view of the indented area. Using a scratch as surface
defect (a), a subcritical crack is grown in the sample, using a
double-bending loading geometry (b). Symmetrical indentation is
then performed on the crack (c). Crack propagation is then
completed under double-bending (d). Two twin sections of the
indented area are then obtained.}\label{protocole tranche}
\end{figure}

\begin{figure}
  \includegraphics[width=16 cm]{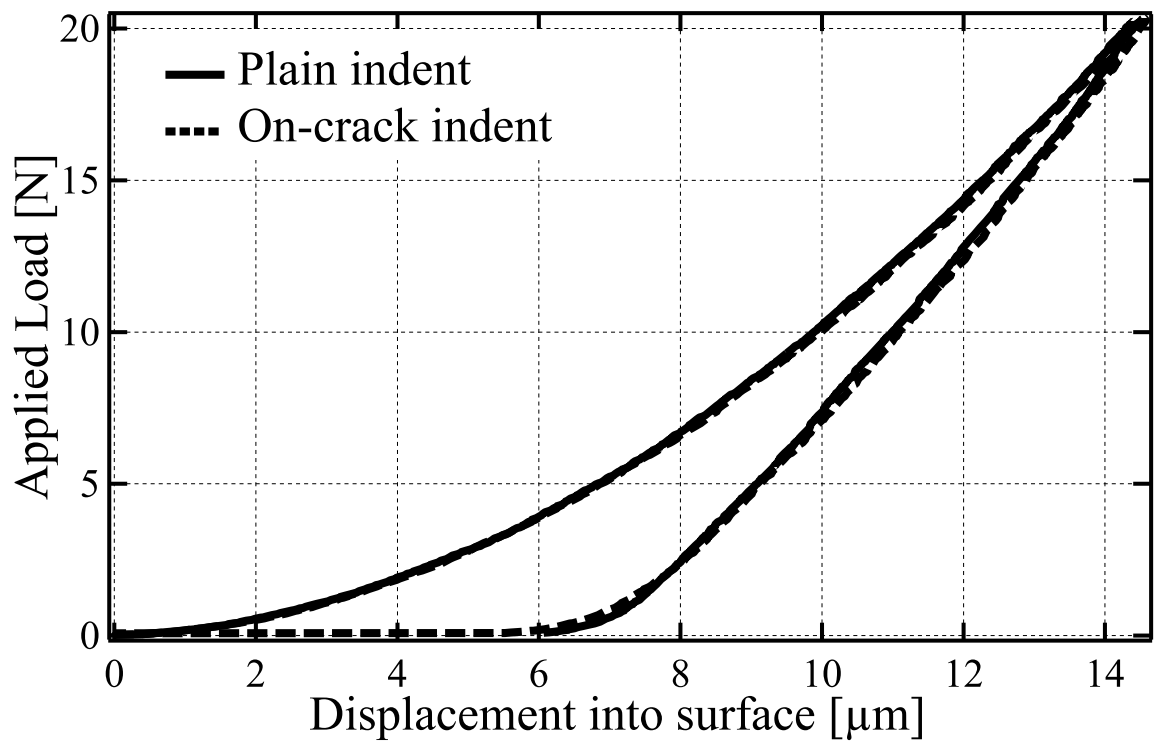}\\
  \caption{Indentation curves obtained for
both on-crack indentation and plain indentation. They almost
superimpose which shows us that the presence of a crack did not
significantly affect the indentation
process.}\label{indentation_curves}
\end{figure}

\begin{figure}
  \includegraphics[width=16 cm]{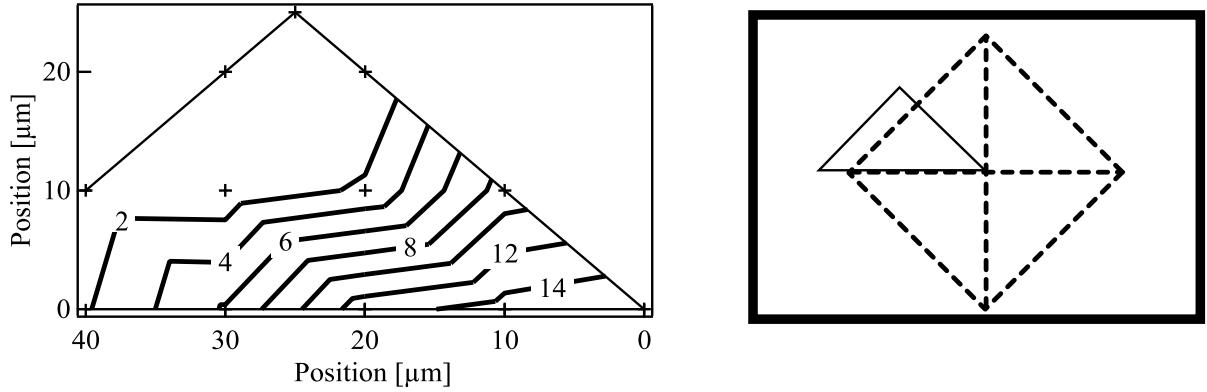}\\
  \caption{Densification map of one eighth of a top
view of an indentation-densified area obtained using  the D$_2$
band position indicator. The center of the indent is at
coordinates (0;0). The bottom thin line figures the edge of the
indent while the right thin line corresponds to the middle of the
pyramid face. The crosses correspond to the measurement points.
The bold lines are the iso-densification lines. Numerical values
are given in percent. Iso-densification lines appear to be
concentric and account for the indenter anisotropy, as
densification appears more important under the indenter edge.
}\label{cartobis}
\end{figure}

\begin{figure}
  \includegraphics[width=16 cm]{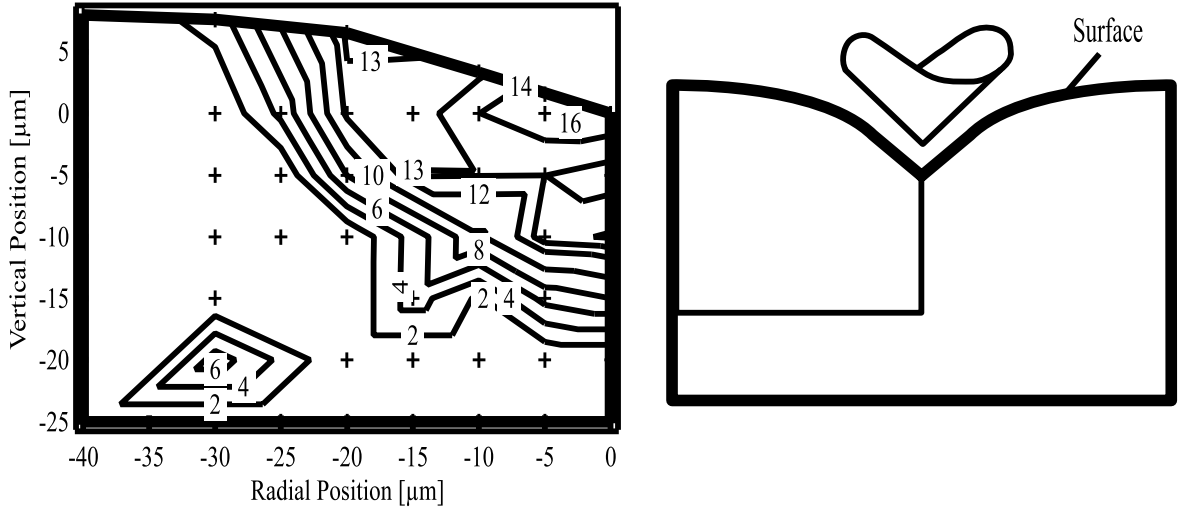}\\
  \caption{Densification map of half a section
view of an indentation-densified area obtained using the D$_2$
band position indicator. The center of the indent is at
coordinates (0;0). The top bold line represents the sample
surface. The crosses correspond to the measurement points. The
black lines are the iso-densification lines. Numerical values are
given in percent. Iso-densification lines appear to be shaped into
concentric bowls in agreement with observations by Hagan et al.
\cite{hagan}}\label{carto}
\end{figure}



\begin{figure}
  \includegraphics[width=8 cm]{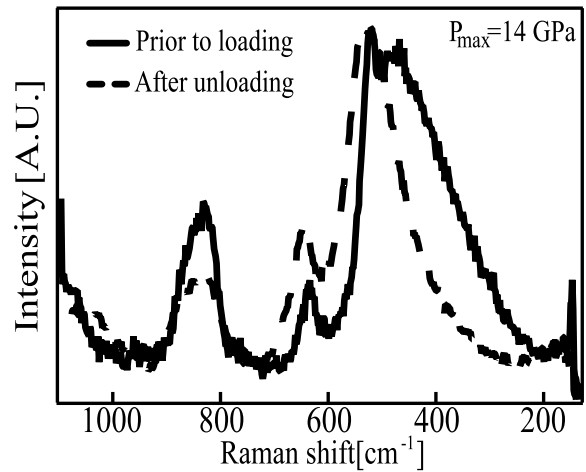}\\
\caption{Raman spectrum obtained with an amorphous silica sample
before (plain line) and after (dotted line) a 14 GPa hydrostatic
loading.}\label{spectra}
\end{figure}

\begin{figure}
  \includegraphics[width=8 cm]{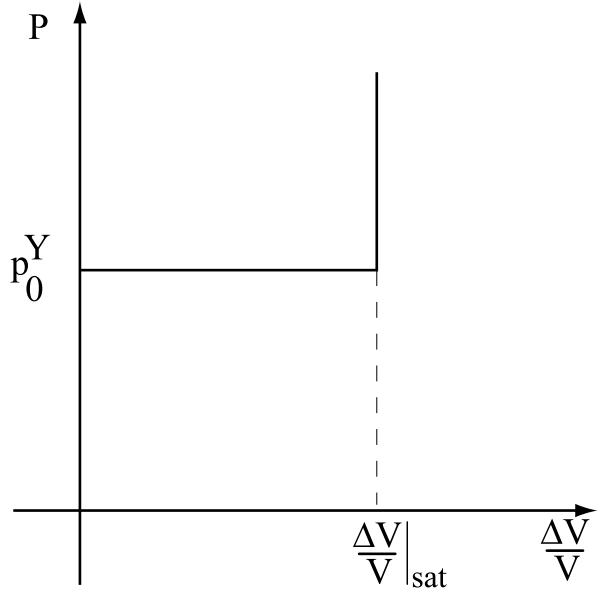}\\
\caption{Expected behavior of a perfect plastic material
under hydrostatic loading. Before pressure reaches the yield
pressure ($p^Y_0=9\mathrm{GPa}$) the sample does not densify. As
its plastic behavior is hardening-free, applied pressure cannot
reach higher values than $p^Y_0$ until irreversible densification,
$\Delta V/V$, reaches its saturation value $\Delta V/V
|_{sat}$.}\label{perfect}
\end{figure}

\begin{figure}
\includegraphics[width=8 cm]{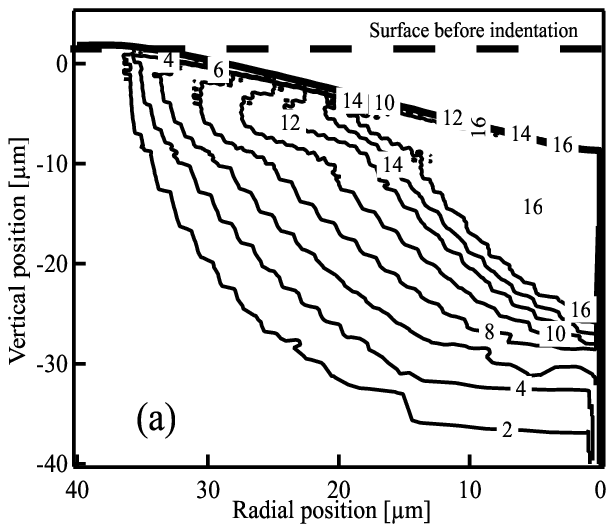} \includegraphics[width=8 cm]{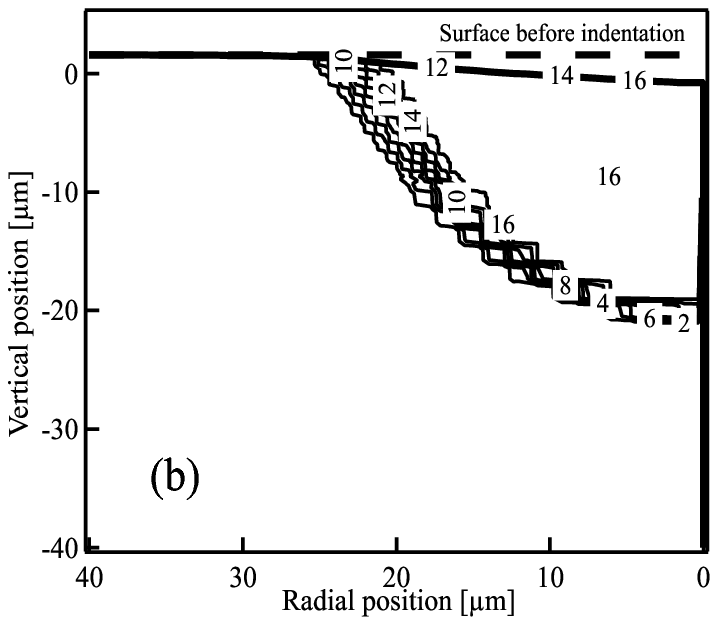}
\caption{Residual densification maps obtained through FE
simulation of a 2D-axisymmetric cone indentation. They were
obtained for two different values of the densification factor, (a)
$\alpha = 0.2$ and (b) $\alpha = 0.6$. Densification values are
given in percent. The results by Xin and Lambropoulos\cite{lamb3d}
are similar to (b)}\label{densif}
\end{figure}

\end{document}